\documentclass{article}
\usepackage{ijcai20}

\usepackage{times}

\usepackage{soul}
\usepackage{url}
\usepackage{enumitem}
\usepackage[hidelinks]{hyperref}
\usepackage[utf8]{inputenc}
\usepackage[small]{caption}
\usepackage{graphicx}
\usepackage{amsmath}
\usepackage{booktabs}

\usepackage{flushend}
\usepackage{multirow}
\usepackage{algorithm,algpseudocode}
\usepackage{algpseudocode}
\usepackage{amsmath}
\usepackage{mathrsfs}
\usepackage{amsfonts}
\usepackage{subfigure}
\title{Learning the Compositional Visual Coherence for\\ Complementary Recommendations}

\author{Zhi Li$^{1}$\and
	Bo Wu$^2$\and
	Qi Liu$^{1,3,}$\footnote{Corresponding Author.}\and
	Likang Wu$^3$\and
	Hongke Zhao$^4$\And
	Tao Mei$^5$\\
	\affiliations
{\normalsize	$^1$Anhui Province Key Laboratory of Big Data Analysis and Application, School of Data Science,\\ University of Science and Technology of China,~~
	$^2$Columbia University~~\\
	$^3$School of Computer Science and Technology, University of Science and Technology of China\\
	$^4$The College of Management and Economics, Tianjin University~~
	$^5$JD AI Research\\}
	\emails
{\normalsize 	\{zhili03, wulk\}@mail.ustc.edu.cn,
	bo.wu@columbia.edu,
	qiliuql@ustc.edu.cn,
	hongke@tju.edu.cn,
	tmei@jd.com}
}

\begin{document}

	\maketitle

	\begin{abstract}

		Complementary recommendations, which aim at providing users product suggestions that are supplementary and compatible with their obtained items, have become a hot topic in both academia and industry in recent years. 
		Existing work mainly focused on modeling the co-purchased relations between two items, but the compositional associations of item collections are largely unexplored.
		Actually, when a user chooses the complementary items for the purchased products, it is intuitive that she will consider the visual semantic coherence (such as color collocations, texture compatibilities) in addition to global impressions.
		Towards this end, in this paper, we propose a novel \textit{Content Attentive Neural Network (CANN)} to model the comprehensive compositional coherence on both global contents and semantic contents. Specifically, we first propose a \textit{Global Coherence Learning} (GCL) module based on multi-heads attention to model the global compositional coherence. Then, we generate the semantic-focal representations from different semantic regions and design a \textit{Focal Coherence Learning} (FCL) module to learn the focal compositional coherence from different semantic-focal representations. Finally, we optimize the CANN in a novel compositional optimization strategy. Extensive experiments on the large-scale real-world data clearly demonstrate the effectiveness of CANN compared with several state-of-the-art methods.
		
	\end{abstract}
	
	\section{Introduction}

	Recommender systems are those techniques that support users in the various decision-making process and catch their interest among the overloaded information. For enhancing user satisfaction and recommendation performances, it is an indispensable part to understand how products relate to each other in recommender systems~\cite{mcauley2015inferring,li18kdd}. Along this line, complementary recommendations~\cite{Yu2019HPBD}, which aim at exploring item compatible associations to enhance the qualities of each item or another, have become a hot topic in both academia and industry in recent years.
	
	\begin{figure}
		\centering
		\includegraphics[width=3.5in]{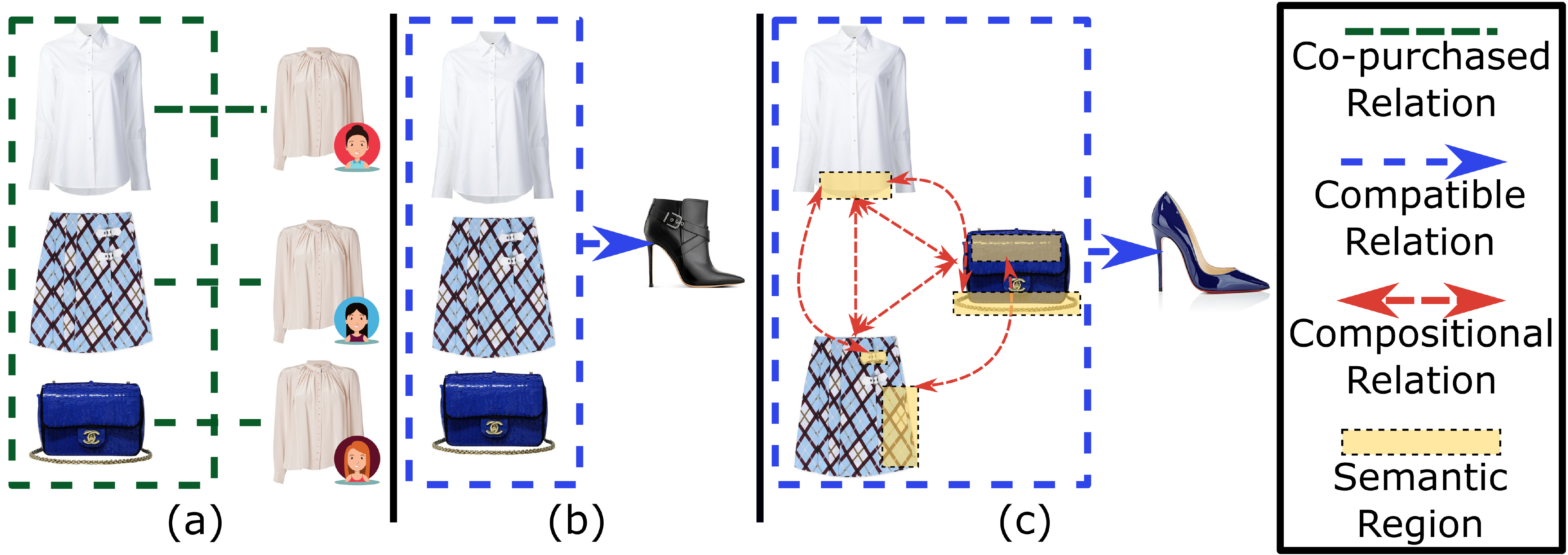}
		\caption{Illustration of complementary recommendations. (a) Recommendations based on co-purchased relations. (b) Recommendations based on compatible relations. (c) Recommendations  based on compositional coherence.}\vspace{-0.2in}
		\label{intro}
	\end{figure}
	
	In the literature, previous researches can be clustered into two groups, i.e., unsupervised methods~\cite{tan2004selecting,zheng2009substitutes} and supervised models~\cite{Zhao2017Improving,he2016learning}. For a long time, researchers mainly used unsupervised methods to model the association rules in the recommendation process~\cite{tan2004selecting}. Meanwhile, some studies proposed supervised approaches to learn complementary relationships, such as co-purchased~\cite{Zhao2017Improving} and content compatibility~\cite{he2016learning}. Recently, many researchers attempted to mine the compatibility of fashion items to better understand the complementary relationships in the clothing recommendations~\cite{han2017learning,hsiao2018creating}. In spite of the importance of existing studies, the exploration of compositional associations in the complementary item recommendations is still limited.

	As a matter of fact, when a user chooses the complementary items for the purchased products, it is intuitive that she will consider the visual coherence (such as color collocations, texture compatibilities) in addition to global impressions. For example, Figure~\ref{intro} shows the case of a complementary recommendation process. If we only consider co-purchased relations, we may recommend a creamy-white shirt to the user as shown in Figure~\ref{intro}(a). That is because the shirt was co-purchased respectively with three query items by other users. When we take consideration of the compatible relations as shown in Figure~\ref{intro}(b), we can find the missing component of the user's purchased collection is a pair of shoes. Then, a pair of black booties may be a suitable recommendation. However, for considering compositional relationships and visual semantic, we can find the mini skirt and shoulder bag are in blue. Therefore, as Figure~\ref{intro}(c) shows, a pair of blue leather pumps is the best-matching. From this example, we can conclude that a good complementary recommender system should model the comprehensive compositional relationships on both global contents and semantic contents. Unfortunately, the exploration of this compositional coherence in the complementary recommendations is still limited.

	In this paper, we provide a focused study of the compositional coherence in item visual contents from two perspectives, i.e., the global visual coherence and the semantic-focal coherence. Along this line, we propose a novel Content Attentive Neural network (CANN) to address the complementary recommendations. More specifically, we first propose a Global Coherence Learning (GCL) module based on multi-heads attention to model the global compositional coherence. Then, considering the importance of visual semantics (such as color, texture), we generate the content semantic-focal representations of color-focal, texture-focal and hybrid-focal contents, respectively. Next, we design a Focal Coherence Learning (FCL) module based on a hierarchical attention model to learn the focal coherence from different semantic-focal representations. Finally, we optimize the CANN in a novel compositional optimization strategy. We conduct extensive experiments on a real-world dataset and the experimental results clearly demonstrate the effectiveness of CANN compared with several state-of-the-art methods.

	\section{Related Works}

	\subsection{Visual Recommendations}
	With the rapid development of deep learning~\cite{zhao2019voice,wu2019estimating} and computer vision, visual recommendations have raised lots of interests in both academia and industry and benefited lots of applications, such as image recommendations~\cite{mcauley2015image}, movie recommendations~\cite{ZhaoLP016} and fashion recommendations~\cite{han2017learning,HouWCLZL19}. Some previous works treated visual recommendations as special content-aware recommendations incorporating the visual appearance of the items~\cite{he2016vbpr}. Along this line, many researchers directly extracted the visual feature by a pre-trained CNN model and enhanced traditional recommender systems, such as Matrix Factorization~\cite{he2016vbpr,HouWCLZL19}. For example, \cite{he2016vbpr} proposed a recommendation framework to incorporate the visual signal of the items. Recently, many researchers utilized deep learning methods to generate item recommendations~\cite{KangFWM17}. To further explore item visual contents, some studies developed deep neural networks to extract the aesthetic information of images~\cite{liu2017deepstyle,YuZHC0Q18}. Although various researches have leveraged visual features in recommendation tasks, they usually treated item visual features as side information and the item visual relations have been largely unexploited. In this paper, we propose the compositional visual coherence, which is learning by the attention-based deep neural networks, to deeply model the item complementary relation for recommender systems.

	\subsection{Complementary Recommendations}

		For enhancing the recommendations, explicitly modeling the complex relations among items under domain-specific applications is an indispensable part~\cite{liu2018illuminating}. Along this line, many researchers focused on exploring the item combination-effect relations, such as substitutable relations~\cite{mcauley2015inferring,mcauley2015image} and complementary relations~\cite{rudolph2016exponential,Yu2019HPBD}. For a long time, many researchers mainly utilized unsupervised learning methods to explore co-occurrence relations for complementary recommendations~\cite{tan2004selecting,zheng2009substitutes}. Recently, more and more studies based on supervised approaches were proposed to model complementary relationships, which were mainly reflected by users' purchases of the complements~\cite{Zhao2017Improving,he2016learning} or item content similarities~\cite{mcauley2015inferring,zhang2018quality}. Along this line, since natural complementary relationships in the fashion items~\cite{chang2017fashion,lo2019dressing}, there was a particular interest in understanding the compatibility~\cite{song2017neurostylist,wang2019outfit} of fashion items to generate complementary recommendations~\cite{han2017learning,hsiao2018creating,Vasileva2018ECCV}. For example,~\cite{he2016learning} learned the complicated and heterogeneous relationships between items and enhance fashion recommendations. \cite{han2017learning} developed Bi-LSTMs to model the outfit completion process and generate the complementary clothing recommendations. Although these works have considered co-reactions between items, the item compositional coherence in the global and semantic contents cannot be depicted and captured well. In this paper, we propose a focal study on exploring the item compositional relationship of visual content to enhance the complementary item recommendations.

	\begin{figure*}
		
		\centering
		\includegraphics[width=7in]{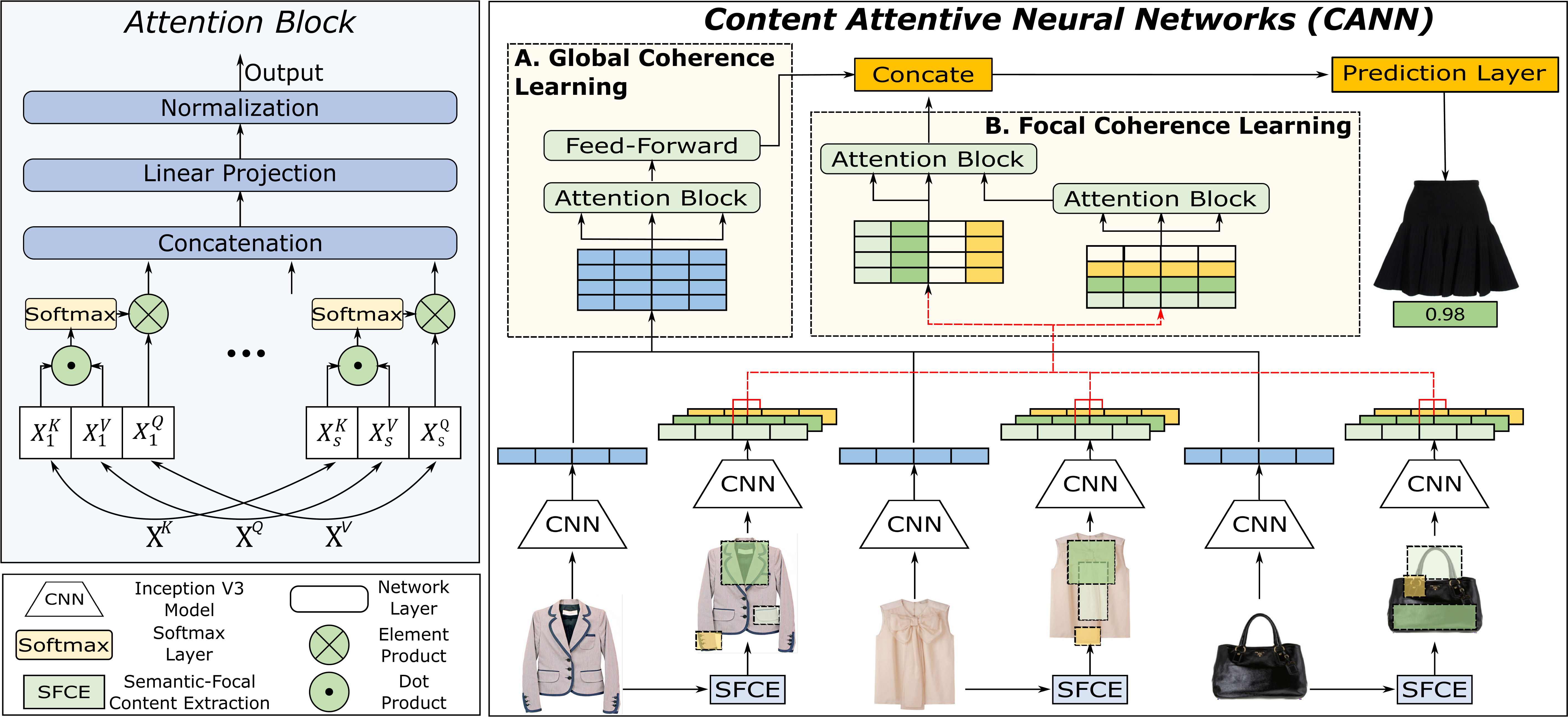}
		\caption{Illustration of Content Attentive Neural network (CANN), which integrates two main components, i.e., A. Global Coherence Learning (GCL) and B. Focal Coherence Learning (FCL).}
		\label{framework}
	\end{figure*}

	\section{CANN: Content Attentive Neural Network}
	In this section, we introduce our proposed framework for addressing the complementary recommendations.

	In the real complementary item choosing process, people always want to buy some items that can be compatible with their purchased products. Suppose we have a set of compatible item collections $S=\{O_1,O_2,...,O_N\}$, and for each collection $O_i=\{p_1,p_2,..,p_k\}$ contains $k$ compatible items $p$. Meanwhile, we have a scenario that a user has purchased a seed collection of items $P=\{p_1, p_2,...,p_i\}$, but she is confused to choose complementary items $P^*$ which can make the item collection compatible. To that end, in this paper, we aim to give a complementary suggestion for the target user to help her make the best-matched choice.

	Along this line, we propose a content-based attention model, i.e., Content Attentive Neural network (CANN), to address the complementation recommendations. As shown in Figure~\ref{framework}, CANN consists of a Global Coherence Learning (GCL) component and a Focal Coherence Learning (FCL) component to jointly learn the compositional relationships from global and semantic-focal contents. In order to simulate users' decision-making process on complementary items, we propose a novel compositional optimization strategy to train our proposed CANN.

	\subsection{Global Coherence Learning}

	Considering the differences from traditional recommendations, the visual contents of items are an important part of the complementary recommendations e.g., the colors should be compatible and styles should be matching. Along this line, we propose an attention-based module, i.e., Global Coherence Learning (GCL) to learn compositional coherence of item global visual contents.
	
	We develop the Inception-V3 CNN model~\cite{szegedy2016rethinking} as the feature extractor to transform item global visual contents (item images) to original feature vectors. Then for each item collection $P=\{p_1, p_2,...,p_k\}$, we can generate the items visual representations $P=\{x_1, x_2,...,x_k\}$ from the item images, where $x_i$ is the feature representation derived from the CNN model for the $i$-th item. We adopt a fully connected layer for all $x_i$ to reduce the feature dimension and make the visual content representation fine-tuned in the model training stage. More formally, we generate the final visual representation as following:
	\begin{equation}
		x_i^{f}=\sigma(x_iW_f^{(1)}+b_f^{(1)})W_f^{(2)}+b_f^{(2)},
	\end{equation}
	where the weight matrices $W_f^{(1)}$, $W_f^{(2)}$ are of shape $\mathbb{R}^{d_c\times d_c}$, $\mathbb{R}^{d_c\times d_f}$, respectively. And $b_f^{(1)}$, $b_f^{(2)}$ are the bias terms, $\sigma$ is the activation function, here, we use \emph{ReLU} function. 
	
	After obtaining the visual representations of the item collection $\widehat{P}=\{x_1^{f}, x_2^{f},... , x_k^{f}\}$, we can generate the item embedding matrix $M\in \mathbb{R}^{k\times d_f}$, where $k$ is the maximum length of the item collections and $d_f$ is the latent dimension of item visual feature. If the item collection length is shorter than $k$, we repeatedly use a `padding' item until the length is $k$. Next, an attention mechanism is applied to model the item compositional coherence in multiple visual spaces. Inspired by the multi-heads attention method~\cite{vaswani2017attention}, in this paper, we use a stacked attention model to capture the global visual coherence. More specifically, we transform the item visual features into various visual spaces and perform a linear projection as $	x_i^{s}=x_iW_s+b_s$, where $x_i^{s}$ is the representation vector of $i$-th item in $s$-th visual space, $W_s$ is the projection matrix of shape $\mathbb{R}^{d_f\times d_s}$, and $b_s$ is the bias term. Then we can model the global coherence between two items in the target collection as:

	\begin{equation}
		\alpha_{i,j}^{s}=\frac{W_s^q{x_i^{s}}\cdot {{(W_s^k x_j^{s})}^T}}{\sqrt{d_s}},\widehat{\alpha}_{i,j}^{s}=\frac{\exp(\alpha_{i,j}^{s})}{\sum_{j=1}^{k}\exp(\alpha_{i,j}^{s})},
		\label{score}
	\end{equation}
	\noindent where $\alpha_{i,j}^{s}$ is the visual coherence score of item $p_i$ and $p_j$, weight matrices $W_s^q$, $W_s^k$ are of shape $\mathbb{R}^{d_s\times d_s}$. Here, we use $\sqrt{d_s}$ to constrain the values of $\alpha_{i,j}^{s}$. Next, the coherence scores are normalized by a softmax function. Then we can generate the item representation $v_i^{s}$ incorporating the global coherence of other items in the $s$-th visual space as:
	\begin{equation}
		v_i^{s}=\sum_{j=1}^{k}{\widehat{\alpha}_{i,j}^{s} \otimes x_j^{s}},
	\end{equation}
	where the $\otimes$ represents the element-wise product. For each visual space, we can generate each item representation including the compositional information. Then, we can fuse all the visual space to generate the global item visual representations, i.e., $v_i=Concate(v_i^{1},v_i^{2},...,v_i^{s})$. Therefore, we model global coherence from multiple visual spaces within an attention-based block and item visual representation $M_{a}=[v_1,v_2,...,v_k]$ is generated. Then, the global coherence learning process can be defined as $M_{a}=f_{a}{(P)}$. However, the global coherence learning process is a linear operation and the interactions between different latent dimensions are largely unexploited. With this in mind, we develop a feedforward network to endow the model with the nonlinear operation:
	\begin{equation}
		f_{n}{(M)}=\sigma(M_{a}W_n^{(1)}+b_n^{(1)})W_n^{(2)}+b_n^{(2)},
	\end{equation}
	where the weight matrices $W_n^{(1)}$, $W_n^{(2)}$ are of shape $\mathbb{R}^{d_f\times d_f}$ and $b_n^{(1)}$, $b_n^{(2)}$ are the bias terms. After that, we generated the attention output from a multi-heads attention block. For modeling the sophisticated compositional coherence of global features, we stack multiple multi-heads attention blocks to build a deeper network architecture. In order to prevent overfitting and unstable training process, we perform the Batch Normalization (BN) layer~\cite{ioffe2015batch} after the output of each attention block. More formally, the output of $b-$th block can be defined as:

	\begin{gather}
		M_{a}^{(b)}=f_{a}({h^{b-1}}) \oplus h^{b-1},\\
		h^{(b)}=BN(W_{bn}f_{n}{(M_{a}^{(b)})})+b_{bn}),
	\end{gather}
	where $\oplus$ is the element plus, $f_{a},f_{n}$ are the attention function and feedforward function, respectively. After the final attention block, we can generate global representations of the seed collection $\mathscr{G}(P)=h^{(b^*)}$, where $h^{(b^*)}$ is the output of the last multi-heads attention block.

	\subsection{Focal Coherence Learning}

	As mentioned above, for learning the compositional coherence from the item content, the content semantic attributes are also important to understanding item characteristics. To that end, we propose a novel Focal Coherence Learning (FCL) to model the compositional coherence from semantic-focal contents, i.e., color-focal contents, texture-focal contents and hybrid-focal contents. Firstly, we develop a Semantic-Focal Content Extraction (SFCE) based on the region-based segmentation methods as regions can yield much richer visual information than pixels~\cite{uijlings2013selective}. Inspired by Selective Search~\cite{uijlings2013selective}, we firstly use a grouping algorithm to merge the semantic regions, which are based on the color or textual similarity computing. Specifically, we first generate the initial semantic regions by a segmentation algorithm~\cite{felzenszwalb2004efficient}. Then we develop the Selective Search algorithm to group regions which are similar in color, texture and both of them, respectively. The similarity can be measured as $Sim({r_i, r_j}) = \sum_{k=1}^{n}{min(C_{i}^{k},C_j^k)}$, where the $C_i$ is the measurement of characteristic feature histogram, i.e., the color histogram and texture histogram, $n$ is the feature dimensionality. More specifically, for each region, we extract three-dimensional color histograms. Each color channel is discretized into 25 bins, and $n = 75$ for the color feature in total. And we obtain $n=240$ dimensional texture histogram by computed SIFT-like~\cite{liu2010exploring} measurements on 8 different orientations of each color channel, for each orientation for each color using 10 bins. Both color histogram and texture histogram are normalized by $L_1$ norm.

	After computing the semantic similarity of regions $i$ and $k$, we can group the similar regions and the target histogram $C_{t}$ can be efficiently propagated by:
	\begin{equation}
		C_{t}=\frac{size(r_i)\cdot C_{i}+size(r_j)\cdot C_{j}}{size(r_i)+size(r_j)}.
	\end{equation}
	
	The size of target regions can be simply computed $size(r_t) = size(r_i)+size(r_j)$. Different from the Selective Search~\cite{uijlings2013selective}, we do not combine the color and texture similarity. Because we want to deeply explore the compositional coherence of different semantic contents. CANN extracts the semantic-focal contents $R_c, R_t, R_h$ respectively based on color similarity, texture similarity and hybrid similarity, and for each semantic-focal content we choose three content regions. With a pre-trained Inception V3~\cite{szegedy2016rethinking}, we can generate the feature vectors $F = [y_c, y_t, y_h]$ of all the semantic-focal contents. Next, we can learn the focal coherence among the content regions.
	
	Actually, considering the computational complexity of the attention mechanism for all the regions in all the items, similar with~\cite{ma2019cdsa}, we propose a hierarchical attention module to model both the semantic-specific and cross-semantic dependency in a distinguishable way. As shown in Figure~\ref{framework}, the input semantic-focal features $V\in \mathbb{R}^{k\times |F|\times |R|\times d_y}$ can be reshaped as input matrices $V_C\in \mathbb{R}^{|F|\times d_C}$ in the semantic-specific space and $V_S\in \mathbb{R}^{k\times d_S}$ in the cross-semantic space. For each seed item collection, $k$ is the seed item numbers, $|F|$ is the number of semantic-focal features, $|R|$ is the number of content regions of each semantic-focal feature, and $d_y$ is the dimension of each region representation. Afterwards, we can compute our focal coherence by hierarchical multi-heads attention as:

	\begin{equation}
		V' = AV =  \hat{A_C}V_S = \hat{A_C}\hat{A_S}V, 	
	\end{equation}
	where $\hat{A_C}$ and $\hat{A_S}$ are attention matrices of semantic-specific and cross-semantic. Similar to GBL, the attention matrices are computed by Eq.\ref{score}. More specifically, we firstly model the compositional coherence of different semantic-focal regions. Then, we learn the attention map of different items. For aligning the attention matrices to the semantic-focal features, we reshape the attention matrices as $\hat{A_C} = A_C \cdot I_C$ and $\hat{A_S} = A_S \cdot I_S$, where $I_C, I_S$ are the identity matrices. To that end, the semantic-focal representation of the item seed collection $P$ can be formulated as $\mathscr{F}(V)=V'$.

	\begin{algorithm}[t]
		\caption{Compositional Optimization Strategy}
					
			\small
			\textbf{Input}: Initialization model $f(P_i,C;\theta)$; 
			The length of the seed collection $k$; 
			The complementary item database $S$; 
			The number of epochs $T$; 
			The size of batch $m$\\
			\textbf{Parameter}: Model parameter $\theta$
			
			\begin{algorithmic}[1]
				\small
			\For{$i=1,2,3,...,T$}
			\State Random sample $m$ seed collections $O\in S$
			\State Initial input mini-batch $Input$ as $\emptyset$
			\For{$O_i~in~Batch$}
			
			\State Random choose an item $p$ from the collection $O_i$
			\State Generate the seed collection $P_i\gets Mask(O_i, p)$
			\If{$|P_i| < k$}
			\State Add an padding to the left of $P_i$ until $|P_i| = k$
			\EndIf
			
			\State Generate the input mini-batch $Input \gets Input \cup P_i$
			
			\EndFor	
			\State Build the training candidates $\mathscr{N} \gets \forall x \in Input$
			\State Update the model $\theta \gets SGD(f(Input,C;\theta),\theta)$

			\EndFor
		\end{algorithmic}
	
	\label{alg}
	\end{algorithm}
	
	\subsection{Optimization Strategy}
	So far, from GCL and FCL modules, we can generate the global and semantic-focal representations, respectively. Next, after a fully connected layer, we can obtain representations of the prediction items. Following~\cite{han2017learning}, we append a softmax layer on the $\widehat{x}$ to calculate the probability of complementary items conditioned on the target seed collections:
	\begin{equation}
		Pr(\widehat{x}|P)=\frac{\exp(\widehat{x}\cdot x_c)}{\sum_{x_c\in \mathscr{N}}{\exp(\widehat{x}\cdot x_c)}},
	\end{equation}
	where $\mathscr{N}$ contains all the items from the seed collections. This can make model learn compositional coherence by means of considering a diverse collection of candidates. For optimizing our CANN, we propose a compositional training strategy to simulate users' decision-making process on complementary items as Algorithm~\ref{alg}. In the training stage, we use the $Mask(O_i, p)$ operation to delete prediction items $p$ from the seed collections $O$. Moreover, we take all items in the mini-batch as the candidate collection $\mathscr{N}$. But in the inference stage, we conduct the candidate collection as the $x_c$. Actually, for some small datasets, we can use all the items as the candidates, but this is not practical for large datasets because of the high dimensional item representations.

	With the compositional optimization strategy, CANN can be trained end-to-end. During training, we minimize the following objective function:
	\begin{equation}
		L(P, \mathscr{N};\theta)=-\frac{1}{|\mathscr{N}|}\sum_{t=1}^{|\mathscr{N}|}{logPr(\widehat{x}|P)},
	\end{equation}
	where $\theta$ denotes the model parameters. Similar with~\cite{han2017learning}, we add a visual-semantic embedding as a regularization. We use \textit{Stochastic Gradient Decent} (SGD)~\cite{robbins1951stochastic} with mini-batch to update them through Algorithm~\ref{alg}.

	\section{Experiments}
	In this section, we first introduce the experimental settings and compared methods. Then, we compare the performance of CANN against the compared approaches on the complementary recommendation task. Then, we make an ablation study on the focal coherence learning process. At last, we conduct a case study to visualize the compositional coherence of our proposed CANN.

	\subsection{Experimental Setups}
	\textbf{Dataset.}
	We evaluate our proposed method on a real-world dataset, i.e., Polyvore dataset~\cite{han2017learning,Vasileva2018ECCV}. It provides the fashion outfits, which are created and uploaded by experienced fashion designers. Indeed, outfit matching is a natural scenario for the content-based complementary recommendations, because the clothes in an outfit are complementary to each other. We use the provided datasets~\cite{han2017learning,Vasileva2018ECCV}, which contain 90,634 outfits. For the reliability of experimental results, we make the necessary specific processing as follows. First, we merge the datasets and remove the noise samples that exclude the seed sets of more than 8 items. It is because a fashion outfit hardly contains more than 8 items in real-world scenarios. Then, we split the dataset into 59,212 outfits with 221,711 fashion items for training, 3,000 outfits for validation and 10,218 outfits for testing. Next, we conduct a testing set (i.e., FITB\_Random) as fill-in-the-blank tasks~\cite{han2017learning}, which we randomly choose candidates as the negative samples from the whole testing set. Meanwhile, in order to further make our testing more difficult and similar to users' decision-making process, we conduct a more specific testing set, i.e., FITB\_Category. In this set, we remove the easily identifiable negative samples and replace these samples with other items which have the same category of the ground-truth.

	\vspace{0.05in}
	\noindent \textbf{Evaluation Metrics.}
	We evaluate our model and all compared methods by two evaluation metrics, i.e., the Accuracy (ACC)~\cite{han2017learning,Vasileva2018ECCV} and Mean Reciprocal Rank (MRR)~\cite{song2017neurostylist,jin2019promotion}.
	
	\vspace{0.05in}
	\noindent \textbf{Implementation Details.}
	We adopt the GoogleNet InceptionV3 model~\cite{szegedy2016rethinking} which was pretrained on ImageNet~\cite{deng2009imagenet} to transform global visual contents (images) and semantic-focal visual contents (regions) to feature vectors. For fair comparisons, we set all the image embedding size of $d_f=512$ unless otherwise noted. The number of visual space is set to $S=4$ and for each visual space, we set $d_s=d_f/S=128$ and $b^*=4$ unless otherwise noted. Our model is trained with an initial learning rate of 0.2 and is decayed by a factor of 2 every 2 epochs. The batch size is set to 9, seed collection length $k$ is set to 8. We stop the training process when the loss on the validation set stabilizes. Our model and all the compared methods are developed and trained on a Linux server with two 2.20 GHz Intel Xeon E5-2650 v4 CPUs and four TITAN Xp GPUs. The datasets and source codes are available in our project pages~\footnote{https://data.bdaa.pro/BDAA\_Fashion/index.html}.

	\subsection{Compared Approaches}
	To demonstrate the effectiveness of CANN, we compare it with the following alternative methods:
	
	\begin{itemize}[leftmargin=*,itemsep=0pt]
		\item \textbf{SetRNN}~\cite{li2017mining}. This method treats the outfit data as a set and develops RNN model to generate the complementary scores of target item sets.
		\item \textbf{SiameseNet}~\cite{veit2015learning}. This model utilizes a Siamese CNN to project two items into a latent space to estimate their similarity. We use an L2 norm to normalize the item visual embedding before calculating the Siamese loss and set the margin parameter to 0.8. 
		\item \textbf{VSE}~\cite{han2017learning}. Visual-Semantic Embedding (VSE) method learns a multimodal item embedding based on images and texts. The resulting embeddings are used to measure the item recommendation scores.
		\item \textbf{Bi-LSTM}~\cite{han2017learning}. This method builds a bidirectional LSTM to predict the complementary item conditioned on previously seen items in both directions. We use the full model by jointly learning the multimodal inputs. 
		\item \textbf{CSN-Best}~\cite{Vasileva2018ECCV}. This method learns a visual content embedding that respects item type, and jointly learns notions of item similarity and compatibility in an end-to-end model. We use the full components with a general embedding size of 512 dimensions in our experiments.
		\item \textbf{NGNN}~\cite{cui2019dressing}. This method learns item complementary relations by the node-wise graph neural networks. This is the state-of-the-art method in fashion compatibility learning and clothing complementary recommendations.
	\end{itemize}
	Besides, for verifying the effectiveness of components, we construct three variant implements based on CANN.
	\begin{itemize}[leftmargin=*,itemsep=0pt]
		\item \textbf{CANN-G}. This is a variant of our model that only uses the Global Coherence Learning (GCL) module to generate complementary recommendations.
		\item \textbf{CANN-F}. This is a specific implementation that only uses Focal Coherence Learning (FCL) module.
		\item \textbf{CANN}. This is the model with all proposed components.
	\end{itemize}

	\subsection{Recommendation Performances}

	We compare CANN with the other methods on the content-based complementary recommendations. The number of candidates is set to 4 for both FITB\_Random and FITB\_Category, which is the same as previous work~\cite{han2017learning,Vasileva2018ECCV,cui2019dressing}. 
	
	The results of all methods on both testing sets are shown in Table~\ref{Tab:performaces}. we can make the following observations: 1) CANN outperforms all the compared methods in both datasets, which indicates the superiority of the proposed model for content-based complementary recommendations. 2) SetRNN and VSE perform the worst on both datasets. That indicates the complementary method is not a trivial problem which can be handled straightly by simple methods. 3) SiameseNet and CSN-Best are similar methods which model the item pair-wise compatibility. These models work better than SetRNN and VSE model, but worse than Bi-LSTM and CANN. This may be due to the fact that SiameseNet and CSN-Best aim to learn the similarity between two fashion items but ignore the compositional process. 4) CANN-F performs worse than CANN-G, that because CANN-F only considers the semantic-focal contents, which may make some information ignored. Moreover, CANN-G outperforms other methods except for CANN. That enables us to safely draw the conclusion that it is advisable to model the compositional coherence of items on both global and semantic-focal contents.

	\subsection{Ablation Study on Focal Coherence}

	To further assess the robustness of the model and necessity of the semantic-focal coherence, we set different semantic contents in FCL module and evaluate the performances as the number of candidate collections increases. We set the FCL module with only color-focal, texture-focal and hybrid-focal contents and compare these models with all our proposed model CANN which contains all the semantic-focal contents.

\begin{figure*}
		\centering
		
		\subfigure[Accuracy on FITB\_Random]{
			\centering
			\includegraphics[width=1.65in]{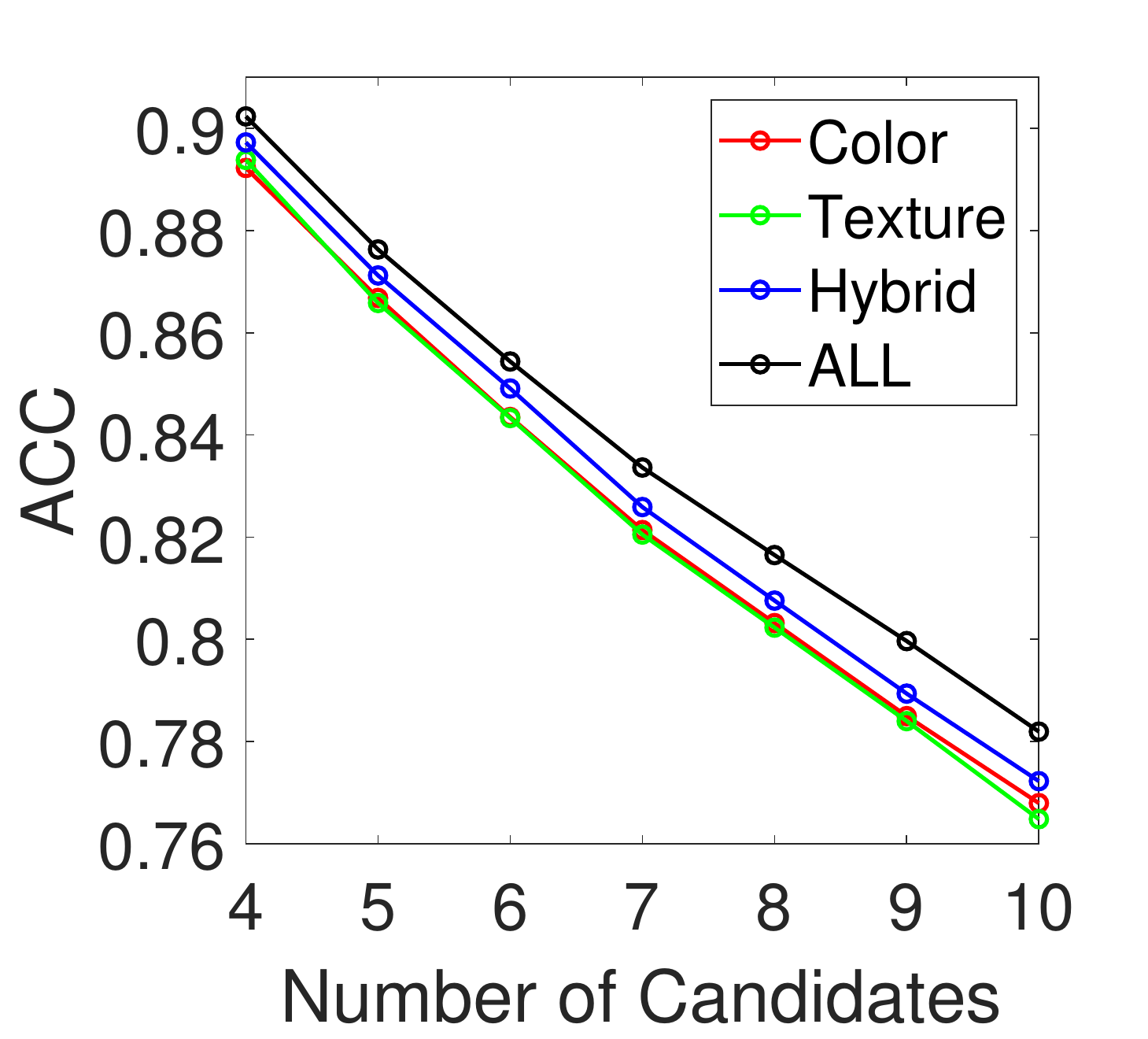}
			
		}
		\subfigure[MRR on FITB\_Random]{
			\centering
			\includegraphics[width=1.65in]{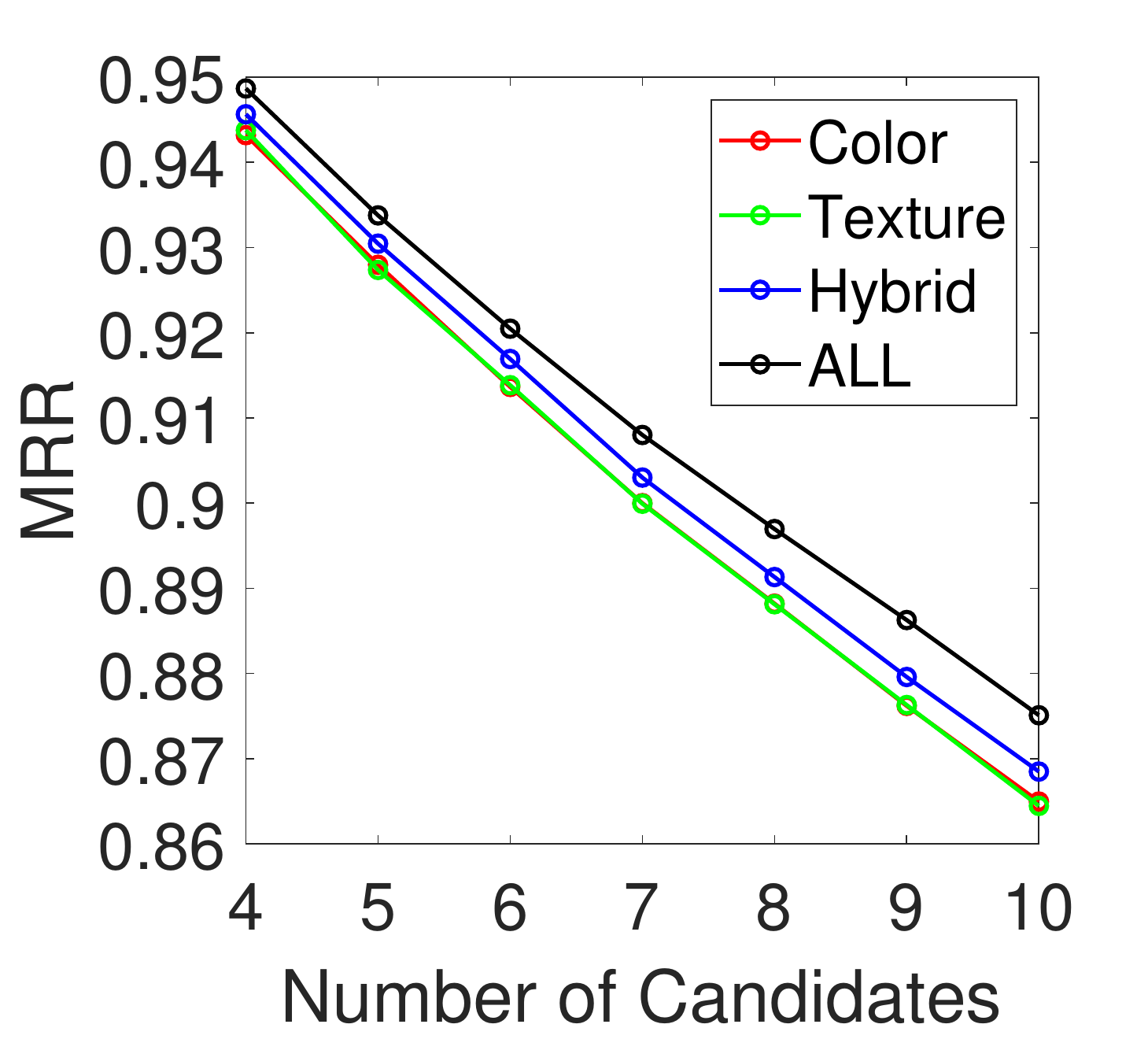}
		}
		\subfigure[Accuracy on FITB\_Category]{
			\centering
			\includegraphics[width=1.65in]{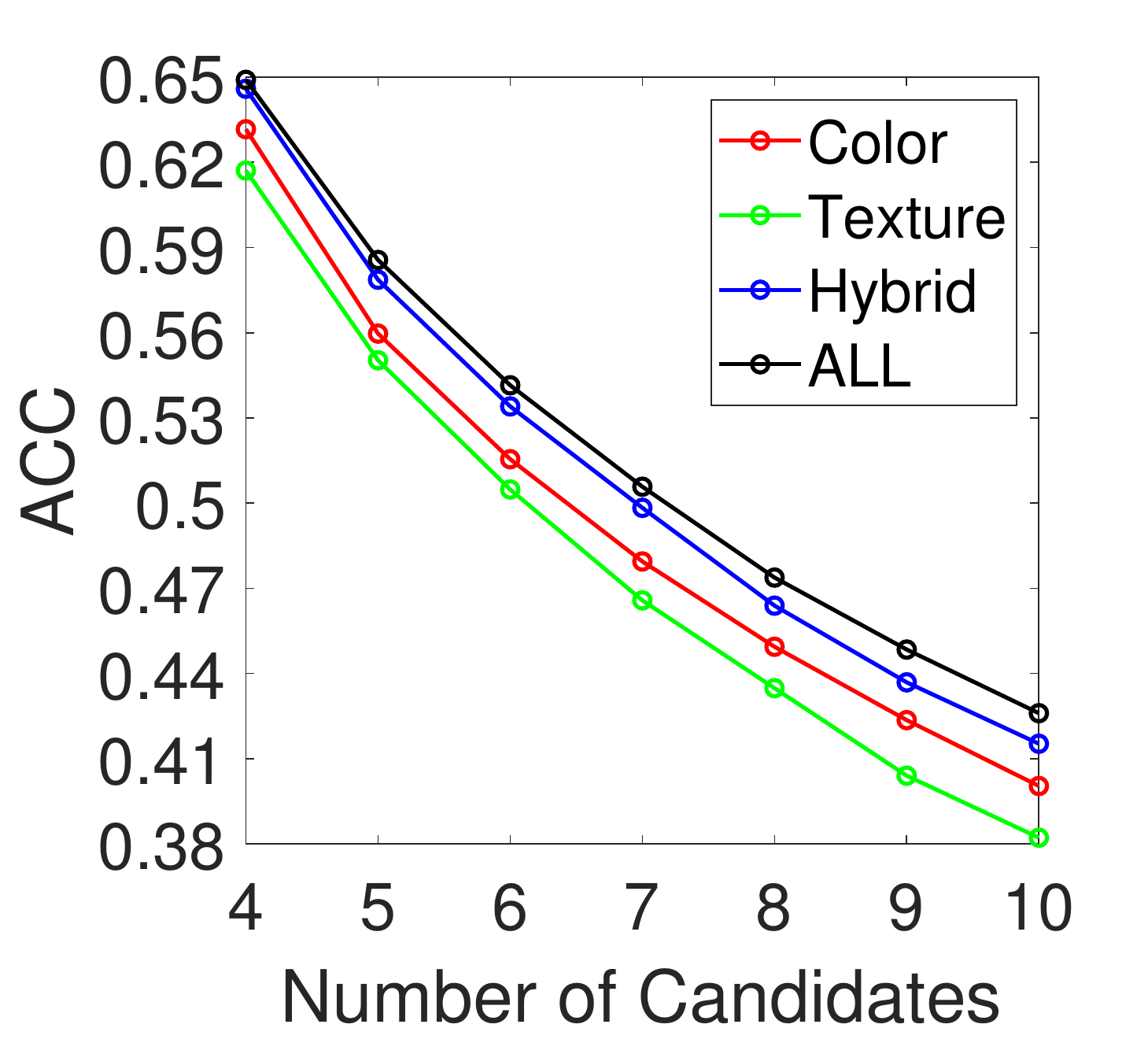}
		}
		\subfigure[MRR on FITB\_Category]{
			\centering
			\includegraphics[width=1.65in]{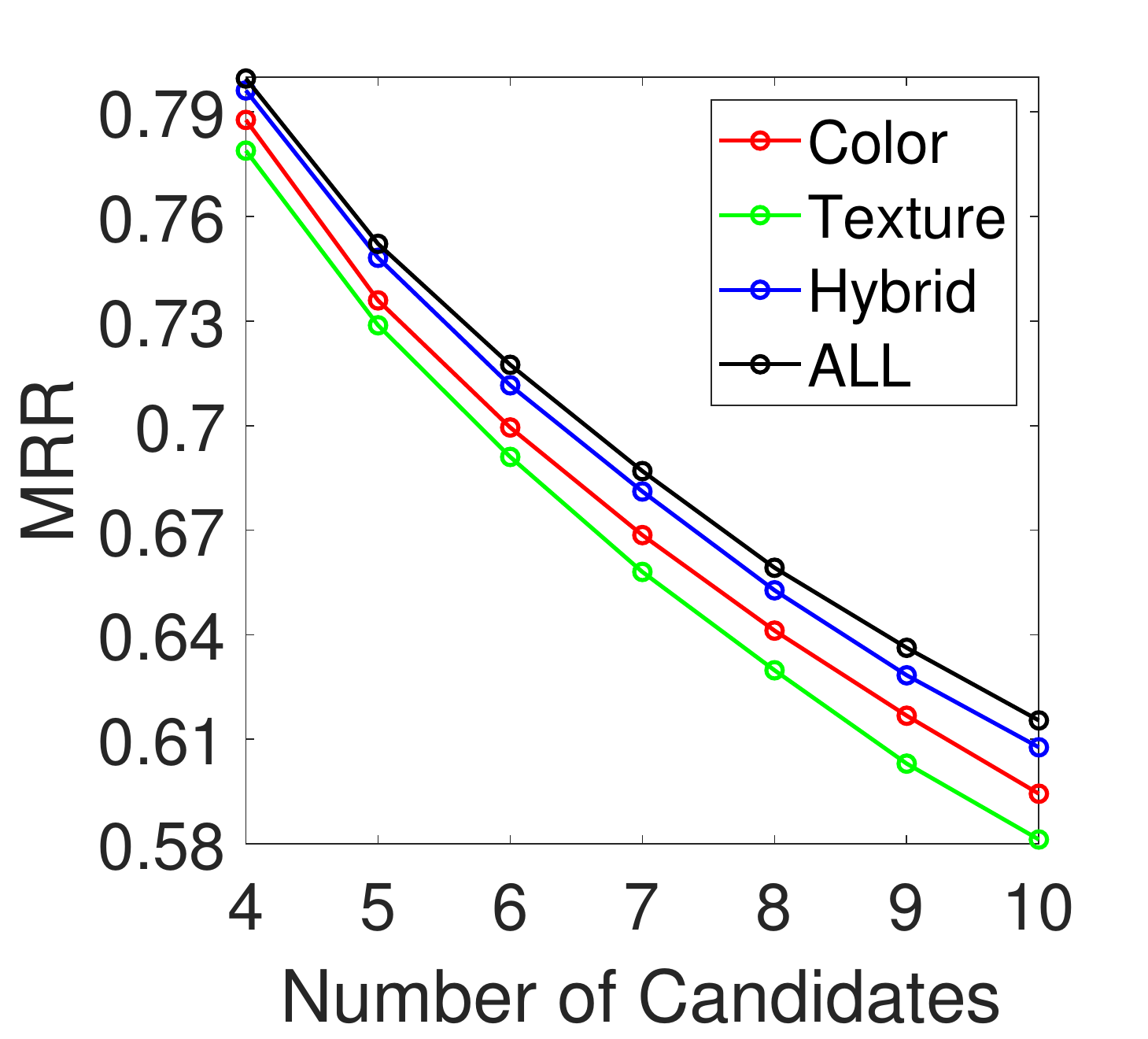}
		}
		\vspace{-0.1in}
		\caption{Results of complementary recommendations over different candidate numbers.}\vspace{-0.05in}
		\label{result}\vspace{-0.05in}
	\end{figure*}

	The results are shown in the Figure~\ref{result}, where the horizontal axis indicates the number of candidate collections. The Color, Texture, Hybrid and ALL represent that CANN uses only color-focal, texture-focal, hybrid-focal and all contents, respectively. From the Figure~\ref{result}, we can get the following observations: 1) CANN with all the semantic-focal contents has outperformed others, which clearly demonstrate the effectiveness of all components in our proposed CANN. 2) The CANN with hybrid-focal coherence learning performs better than color-focal and texture-focal methods. Meanwhile, the color-focal model outperforms the texture-focal model. These observations imply that the color is an important factor for content-based complementary item recommendations. 3) CANN with all semantic-focal contents outperforms other single semantic-focal models on FITB\_Category with a larger margin than on FITB\_Random. These observations imply that semantic-focal contents can help the model to better understand the item compositional relationships and generate the best-matched complementary item suggestions.
			\begin{table}[t]
				\renewcommand\arraystretch{1.2}
		\caption{Performance comparisons of CANN with alternative methods on two specific testing sets.}	\vspace{-0.05in}
		\begin{tabular}{ccccccc}
			\hline
			
			\multirow{2}*{Approaches}& \multicolumn{2}{c}{FITB\_Random}& \multicolumn{2}{c}{FITB\_Category}\\
			\cline{2-3}
			\cline{4-5}
			
			\multirow{2}*{}& Accuracy& MRR&Accuracy& MRR\\
			
			\hline
			SetRNN 		    &29.6\%& 48.1\%& 28.7\%& 46.1\% \\
			SiameseNet	&52.2\%& 71.6\%& 54.0\%& 72.8\% \\
			
			VSE			&29.2\%& 49.1\%& 30.2\%& 53.2\% \\
			Bi-LSTM		&83.6\%& 91.1\%& 58.2\%& 75.7\% \\
			
			CSN-Best	&58.9\%& 76.1\%& 56.1\%& 74.2\% \\
			NGNN    	&87.3\%& 93.2\%& 57.3\%& 74.9\% \\
			\hline
			CANN-G		&{88.8\%}& { 94.1\%}& {62.4\%}& {78.1\%} \\
			CANN-F		&{71.9\%}& { 84.1\%}& {56.7\%}& {74.7\%}\\
			CANN 		&\textbf{90.7\%}& \textbf{ 95.1\%}& \textbf{66.5\%}& \textbf{80.9\%}\\
			\hline
		\end{tabular}\vspace{-0.05in}
		\label{Tab:performaces}
	\end{table}

	\subsection{Visualization of the Attention Mechanism}

\begin{figure}
		\centering 
		\includegraphics[width=3.4in]{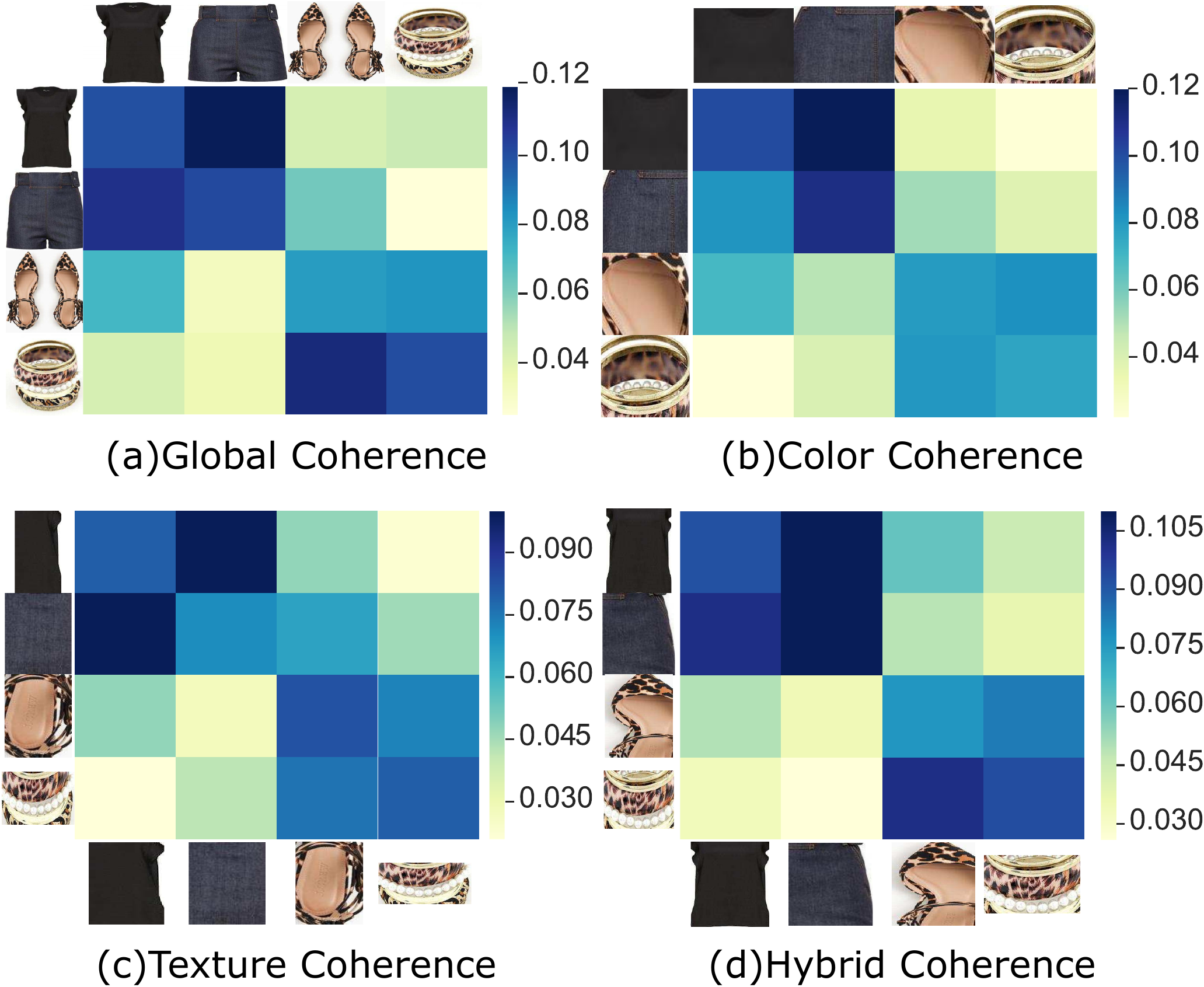}		
		\caption{Visualization of the compositional coherence scores between two items in both of the global and semantic-focal contents.}\vspace{-0.15in}
		\label{Weights}
	\end{figure}

	To further illustrate the learning and expression of compositional visual coherence in our model, we visualize the intermediate results of the coherence score $\widehat{\alpha}_{i,j}$ in Eq.~\ref{score}. Figure~\ref{Weights} illustrates the item compositional relations in an example. For the better visualization, we select one semantic region generated by our model for each semantic-focal content. The color in Figure~\ref{Weights} changes from light to dark while the value of coherence score increases.
	
	From Figure~\ref{Weights}, it is worth noting that the coherence scores between the t-shirt and shorts are higher than others in all coherence spaces. Meanwhile, the scores between shoes and bracelet are also quite high. That is intuitive that the black t-shirt and dark blue shorts are similar in color. The shoes and bracelet are similar in leopard print style. These observations imply that our proposed CANN can provide a good way to capture the visual coherence for the complementary items from both global and semantic-focal views.	
	
	\section{Conclusion}
	In this paper, we proposed a novel framework, the Content Attentive Neural Network (CANN), to address the problem of content-based complementary recommendations. 
	For generating complementary item recommendations, the global and semantic contents (such as color collocations, texture compatibilities) are indispensable parts to understand the comprehensive compositional relationship among items. Along this line, we provided a focused study on the compositional coherence in item visual contents. More specifically, we first proposed a Global Coherence Learning (GCL) module based on multi-heads attention to model the global compositional coherence. Then, we generated the content semantic-focal representations and designed a hierarchical attention module, i.e., Focal Coherence Learning (FCL), to learn the focal coherence from different semantic-focal contents. Next, for simulating users' decision-making process on complementary items, we optimized the CANN in a novel compositional optimization strategy. Finally, we conducted extensive experiments on a real-world dataset and the experimental results clearly demonstrated the effectiveness of CANN compared with several state-of-the-art methods.

	In the future, we would like to consider multi-modal information, such as item descriptions and categories for the deep exploration of the item complementary relationships. Moreover, we are also willing to investigate the domain knowledge about aesthetics assessment and make the recommender systems more explainable.

	\section*{Acknowledgments}
	
	This research was supported by grants from the National Natural Science Foundation of China (Grants No. 61922073, 61672483, U1605251). Qi Liu acknowledges the support of the Youth Innovation Promotion Association of CAS (No. 2014299) and the USTC-JD joint lab. Bo Wu thanks the support of JD AI Research.
	We special thanks to all the first-line healthcare providers, physicians and nurses that are fighting the war of COVID-19 against time.

	\bibliographystyle{named}
	\bibliography{ijcai20}
\end{document}